# The Poisson Boltzmann equation and the charge separation phenomenon at the silica-water interface: A holistic approach

*Maijia Liao[1], Li Wan[2], Shixin Xu[3], Chun Liu[4], Ping Sheng[1,*]*

[1]*Department of Physics, Hong Kong University of Science and Technology Clear Water Bay, Kowloon, Hong Kong, China*
[2]*Department of Physics, Wenzhou University, Zhejiang Province, China*
[3] *School of Mathematical Sciences, Soochow University, Suzhou, China*
[4] *Department of Mathematics, Pennsylvania State University, University Park, Pennsylvania 16802,USA*

## Abstract

The Poisson-Boltzmann (PB) equation is well known for its success in describing the Debye layer that arises from the charge separation phenomenon at the silica-water interface. However, by treating only the mobile ionic charges in the liquid, the PB equation essentially accounts for only half of the electrical double layer, with the other half—the surface charge layer—being beyond the PB equation's computational domain. In this work, we take a holistic approach to the charge separation phenomenon at the silica-water interface by treating, within a single computational domain, the electrical double layer that comprises both the mobile ions in the liquid and the surface charge density. The Poisson-Nernst-Planck (PNP) equations are used as the rigorous basis for our methodology. This holistic approach has the inherent advantage of being able to predict surface charge variations that arise either from the addition of salt and acid to the liquid, or from the decrease of the liquid channel width to below twice the Debye length. The latter is usually known as the charge regulation phenomenon. We enumerate the "difficulty" of the holistic approach that leads to the introduction of a surface potential trap as the single physical input to drive the charge separation within the computational domain. As the electrical double layer must be overall neutral, we use this constraint to derive both the form of the static limit of the PNP equations, as well as a global chemical potential $\mu$ that is shown to replace the classical zeta potential (with a minus sign) as the boundary value for the PB equation, which can be re-derived from our formalism. In contrast to the zeta potential, however, $\mu$ is a calculated quantity whose value contains information about the surface charge density, salt concentration, etc. By using the Smoulochowski velocity, we define a generalized zeta potential that can better reflect the electrokinetic activity in nano-sized liquid channels. We also present several predictions of our theory that are beyond the framework of the PB equation alone—(1) the surface capacitance and the so-called pK and pL values that reflects the surface reactivity, (2) the isoelectronic point at which the surface charge layer is neutralized, in conjunction with the surface charge variation as a function of the solution acidity (pH), and (3) the appearance of a Donnan potential that arises from the formation of an electrical double layer at the inlet regions of a nano-channel connected to the bulk reservoir. All theory predictions are shown to be in good agreement with the experimental observations.

*Correspondence email: sheng@ust.hk

# I. Introduction

*1a. Physical motivation*

Charge separation at the liquid-solid interface, and the subsequent formation of an interfacial electrical double layer, is responsible for a variety of phenomena that are collectively known as "electrokinetics." As the physical basis for motivating this work, we shall focus our attention on the silica-water interface. The silica surface can have either dangling *Si* bonds or dangling *Si-O* bonds. When the silica surface comes into contact with water, one neutral water molecule can dissociate into an $OH^-$ ion and an $H^+$ ion, which would combine, respectively, with *Si* and *Si-O* to form two silanol (*SiOH*) groups. The silanol group is understood to be unstable in an aqueous environment and can easily lose (or gain) a proton. The dissociated protons must stay in the neighborhood of the interface owing to the electrostatic interaction with the negative charges left on the interface. In this manner an electrical double layer (EDL) is established. EDL is characterized by the presence of high concentrations of excess mobile charges in the liquid, required to shield the surface net charges, which are fixed. When an electric field is applied to systems that display charge separation at the liquid-solid interface, electrokinetic phenomena invariably arise. These can be, for example, electroosmosis in which the application of an electric field tangential to the charge separation interface induces liquid flow, or electrophoresis in which a particle with an electrical double layer at its surface would move at a fixed speed under the application of an external electric field.

The surface charge layer, which constitutes half of the electrical double layer, is known to react to the condition of the liquid solution. In particular, with the addition of acid (i.e., excess $H^+$ ions), it has been experimentally observed that the surface charge layer can be neutralized. We define the pH value, which characterizes the proton concentration, as $-\log_{10}[H^+]$. Here the square brackets denote the concentration of the denoted ion species; hence a low pH implies a high concentration of $H^+$ ions. The pH value of the aqueous solution where the surface net charge density is zero, is defined to be the isoelectric point (IEP). When the pH further decreases below the IEP, the net surface charge can be observed to change sign [1,2]. The whole process may be described by the following reaction at the interface [3-5]:

$$SiO^-(\text{pH>IEP}) \xleftrightarrow{H^+} SiOH(\text{pH=IEP}) \xleftrightarrow{H^+} SiOH_2^+(\text{pH<IEP}).$$

Many physical and chemical properties of water/solid oxide interfaces are linked to the phenomenon of IEP [6,7] such as competitive adsorption, interface distribution of ions and surface hydration [1,2]. Thus, it is well-established that the surface charge can be affected by the ionic concentrations in the liquid. This phenomenon is generally denoted as "charge regulation" [8]. Besides the IEP, physical properties of nanofluid channels have also been observed to deviate from the bulk. Here we mention only two such nano-channel phenomena: the charge regulation behavior in which the net surface charge is observed to continuously decrease as a function of the channel width, and the appearance of a so-called Donnan potential which characterizes the electrical potential difference between the inside of a nano fluid channel and the bulk reservoir to which it is attached. Donnan potential vanishes for bulk channels and increases with decreasing channel width.

*1b.    The Poisson-Boltzmann equation and its computational domain*

Classical mathematical treatment of the charge separation phenomenon at the liquid-solid interface is centered on the Poisson-Boltzmann equation:

$$\nabla^2 \bar{\varphi} = \frac{1}{\lambda_D^2} \sinh(\bar{\varphi}) \ , \tag{1}$$

where $\bar{\varphi} = e\varphi/k_B T$, $\varphi$ denotes the electrical potential, $e$ the electronic charge, $k_B$ the Boltzmann constant, and $T$=300 K denotes room temperature.  Here we have assumed all the ions to be monovalent in character, $\lambda_D = \sqrt{\varepsilon k_B T/(2e^2 n^\infty)}$ is the Debye length, where $\varepsilon$ denotes the dielectric constant of the liquid, and $n^\infty$ being the bulk ion density, which must be the same for the positive and negative ions. Equation (1) is usually solved by specifying a boundary value, $\zeta$, denoted the zeta potential, at the interface between the surface charge layer and the screening layer in the liquid.  The formulation of the PB equation represents a historical breakthrough in the mathematical treatment of the charge separation phenomenon. Its accurate prediction of the Debye layer has withstood the test of time and many experiments.

In what follows it is necessary to specify the geometric shape of the liquid channel.  For simplicity, we shall use cylindrical channel with radius *a* in our considerations.  Exception will be noted.  It should be emphasized, however, that although in the present work the cylindrical geometry is used to ensure consistency, the general underlying approach is not particular to any given geometry of the liquid channel.

It should be noticed that the right hand side of Eq. (1) is a monotonic function that denotes the net charge density.  There is an absolute zero potential value associated with the PB equation's right hand side that specifies the point of zero net charge density.   For a large enough liquid channel, the center of the channel must be neutral.  Hence we can associate the zero potential with the center of our (large) cylindrical channel. It follows that the integration of the right hand side of Eq. (1), from center to the boundary (where there is a non-zero $\zeta$ ) must lead to a net nonzero charge.  This is precisely the net charge in the Debye layer, which must be compensated by the surface charge lying just beyond the computational domain of the PB equation.

From the above brief description it becomes clear that notwithstanding its historical achievement, the PB equation describes only half of the electrical double layer—the liquid half that comprises the mobile ions.  As mentioned previously, the addition of salt and/or acid to liquid, or the variation of the liquid channel width, can affect the surface charge layer, which constitutes the other half of the electrical double layer, and consequently the zeta potential value that serves as the boundary value to Eq. (1).  Such variations are beyond the PB equation framework alone and hence their explanations require additional theoretical and/or experimental inputs, in the form of phenomenological parameters and equations that must be incorporated [9-15] and then handled in conjunction with the PB equation. The traditional approaches, which invariably start with the mobile ions in the fluid (accounted for by the PB equation) and the surface charge density as two separate components of the problem, would link the two by using a surface reaction constant, the so-called pK (or pL) value (defined below in Section 6a) and the electrical potential value at the liquid-solid interface [9,11,13,16]. Overall charge neutrality is then reflected in the consistent solution of the electrical potential of the problem, and charge regulation phenomenon can be accounted for in this manner. However, it should be noted that the pK and pL values are experimental inputs which can take somewhat different values in different pH ranges. Alternatively, the problem can also be cast in the form of a free energy of the system, with postulated attractive potentials at the solid surface, each for a particular ionic species. The surface charge densities that result are then coupled by using Lagrange multipliers to an ionic reservoir of a given ionic strength [17, 18]. The Lagrange multipliers are interpreted as chemical potentials for the different ionic species.

*1c.     Features of the holistic approach*

In view of the above, an obvious question arises: Can there be a holistic approach in which the Debye layer and the surface charge layer are treated within a unified framework from the start, *using a single computational domain*? It is the purpose of this work to answer this rhetorical question in the affirmative. In particular, the holistic approach should have the following three features. (1) All ionic densities, including that for the surface charge density, should appear on the right hand side of the Poisson equation. This would ensure all the electrical interactions be accounted for in a consistent manner, including those between the surface charge density and the ions in the Debye layer, the interaction between all the ions within the Debye layer, and the interaction between the all the ions within the surface charge density. A direct implication is that the spatial integral of the right hand side of the Poisson equation must be zero. This feature represents a fundamental departure from the traditional PB equation. (2) Within the above context a charge separation mechanism, based on energy consideration, should be introduced to drive the formation of the surface charge layer. (3) The form of the PB equation must be re-derivable within the reduced domain, i.e., within the traditional PB equation domain that excludes the surface charge layer.

A particular advantage of the holistic approach, as compared to the traditional approach, lies in the computational simplicity for complex interfacial geometries, in which the surface charge density can vary along the interface. A simple example along this direction is given in Section 6c, in which the appearance of the Donnan potential in a finite nanochannel is delineated by a detailed map of the ionic charge density variation at the inlets of the nanochannel.

The starting point of our approach is the Poisson-Nernst-Planck (PNP) equations, which accurately describe the electrical interaction between the ionic charges and their diffusive dynamics. Since the electrical double layer must be overall charge-neutral, this condition will be used to advantage in deriving the relevant equations and a global chemical potential. The PB equation can be re-derived in our formalism (within a reduced computational domain), but with a new clarification for the meaning of the zeta potential that was traditionally treated as the boundary condition for Eq. (1).

*1d.   Outline of the paper*

In order to make the present manuscript self-contained, it is necessary to include materials that have been previously appeared in ref. [19]. However, in the present work the mathematical approach contains an important new element (see Section 4c) that enabled all the new predictions presented in Section VI.

In what follows, Section II introduces the PNP equations and their boundary conditions. The "difficulty" of the holistic approach, which can be stated as the absence of charge separation in an overall charge-neutral domain by applying uniform boundary conditions, is briefly described. That leads naturally to the introduction of a surface potential trap in Section III that serves as the physical input to drive the charge separation process, in conjunction with the formation of the surface charge density. In Section IV we describe the derivation of the charge-conserved Poisson-Boltzmann (CCPB) equation, followed by an enumeration of the elements in the mathematical formulation of the approach. In Section V we re-write the CCPB equation in conjunction with the definition of a global chemical potential $\mu$, followed by a re-derivation of the PB equation and a description of the solution approach. Section VI presents some predictions of our holistic approach. In Section VII we conclude with a brief summary.

## II.  The Poisson-Nernst-Planck equations

*2a.   Equations expressing charge conservation and electrical interaction*

In an overall charge-neutral fluid with a given density of positive ions $n_+(\boldsymbol{x})$ and negative ions $n_-(\boldsymbol{x})$, where $\boldsymbol{x}$ denotes the spatial coordinate, the spatial average of both $n_+$ and $n_-$ must be the same, denoted by $n^o$. In anticipation of later developments, we want to note here that in our holistic approach, $n^O$ comprises both the bulk ion density, $n^\infty$, and the interface-dissociated charge density, $\sigma$ (see Section IV, Eq. (10)).

The dynamics of the ions and their interaction should satisfy the charge continuity equation and the Poisson equation. This is expressed in a rigorous manner by the Poisson-Nernst-Planck (PNP) equations [20-25]:

$$\frac{dn_-}{dt} + \nabla \bullet \boldsymbol{J}_- = 0, \qquad (2a)$$

$$\frac{dn_+}{dt} + \nabla \bullet \boldsymbol{J}_+ = 0, \qquad (2b)$$

$$\boldsymbol{J}_\pm = -D\left(\nabla n_\pm \pm \frac{e}{k_B T} n_\pm \nabla \varphi\right), \qquad (2c)$$

$$\nabla^2 \varphi = -\frac{e(n_+ - n_-)}{\varepsilon}. \qquad (2d)$$

Here all the ions are taken to be monovalent, $D$ is the diffusion coefficient for negative and positive ions, here assumed to be the same for both species, $\boldsymbol{J}_\pm$ denotes the ion flux for either the positive or the negative ions; they are seen to comprise the sum of two terms: one for the diffusive flux and the other for the drift (or convective) flux. Both components are seen to be the spatial derivatives of the local chemical potentials, i.e., the ion concentration and electrical potential. These components of the chemical potential are especially noted in order to distinguish them from the global chemical potential that expresses the overall charge neutrality condition, presented in Section V. Equations (2a)-(2c) describe the charge continuity condition for both the positive and negative ions, while Eq. (2d) is the Poisson equation relating the net ion charge density to the electrical potential $\varphi$. The PNP equations can be solved numerically; an analytical solution to the one dimensional PNP equations was proposed only recently [26-29]. The PNP equations were used to study ion transport dynamics [30-34]; here they are regarded as the basis of our holistic approach.

In this work we choose to treat the simplified problem in which the system is overall electrically neutral, with ions represented by point particles each carrying a single electronic charge, with no chemical distinctions. An exception is made with respect to the distinction between the ions that can participate in the surface-specific adsorption at the fluid-solid interface and the non-surface-specific ions. The latter refers to those salt ions which do not interact or adsorb onto the fluid-solid interface (Section IV). The conditions of electrical neutrality and constant (average) ion density are noted to be compatible with the PNP equations and the relevant boundary conditions, given below.

*2b.     Boundary conditions and the computational domain*

The kinematic boundary conditions for the PNP equations may be easily stated as follows. At the liquid-solid interface, we should have $\boldsymbol{J}_\pm \bullet \hat{\boldsymbol{n}} = 0$, where $\hat{\boldsymbol{n}}$ denotes the interfacial unit normal. These conditions guarantee the conservation of $n_\pm$ and hence the overall charge neutrality if the system starts out to be neutral.

The electrical boundary conditions at the liquid-solid interface are the most important since they give rise to the EDL and hence the electrokinetic phenomena. Traditionally this can be either the Dirichlet type boundary condition in which a constant potential is specified, or a Neumann type boundary condition in which a constant normal electric field is given. However, we shall see that neither can yield charge separation within a computational domain that is overall charge neutral.

For clarity, in Fig. 1 we draw the liquid channel geometry to be considered below. Exception will be noted (see Section VI). If a cylindrical channel is sufficiently long as compared to its cross sectional dimension, then any effects introduced by its two ends can be ignored. A simple way to represent this geometry is a very large doughnut as shown in Fig. 1 in which the two ends of the cylindrical channel are closed to form a loop. If we consider any arbitrary cross section of the large doughnut as indicated by the shaded area in Fig. 1, then from symmetry consideration such a cross section must also be charge neutral. Let us consider such a cross section as our computational domain. This is consistent with our intent to consider the electrical double layer as a whole, so that there is overall charge neutrality.

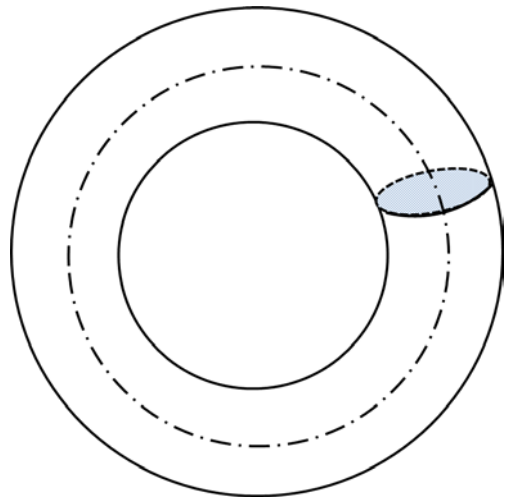

**Figure 1. Doughnut geometry formed by cylindrical channel without end effect. If the system is overall charge neutral, then from symmetry consideration any arbitrarily selected shaded cross section must also be charge neutral.**

Since in the PNP equations the electrical potential appears only in the form of its spatial derivatives, hence the solution to the PNP equations must be insensitive to any additive constant potential. It follows that any constant potential boundary condition should be the same as any other. In particular, we can use the zero potential boundary condition, which would yield trivial solution in view of the fact that there is nothing in the computational domain to break the symmetry between the positive and negative charges. As to the Neumann boundary condition, it follows from the Gauss theorem that the only physically compatible Neumann boundary condition is zero normal electric field, which would also yield trivial solutions. Therefore, for the overall neutral computational domain, uniform boundary condition is not possible to describe the physical situation. This conclusion, which may be denoted as the "difficulty" of the holistic approach, can be easily verified by using the static limit of the PNP equations, i.e., the charge-conserved Poisson Boltzmann equation, given in Section IV.

In what follows, we will use the zero potential boundary condition, but with a mechanism inside the computational domain to drive the charge separation and the consequent formation of the surface charge density.

## III.     Surface potential trap

*3a.     Energetics of interfacial charge separation*

We propose a (charge-neutral) surface potential trap model at the fluid-solid interface to serve as the physical input for driving the interfacial charge separation, attendant with the formation of a surface charge layer. To motivate this model, let us consider the silanol group at the water-silica interface. The depth of the surface potential trap is intended to be indicative of the free energy relevant to the charge dissociation process. In other words, we attribute a constant free energy cost to each ion pair ($SiO^-$ and $H^+$) generated. It is essential to note that the ions in the potential trap are $SiO^-$ formed by $SiOH + OH^- \Leftrightarrow SiO^- + H_2O$. Hence in place of the $SiO^-$, we will use $OH^-$ instead. In what follows, we will associate the surface potential trap only with the $OH^-$ and $H^+$ ions by excluding, via mathematical means, those salt ions that do not physically react with the silica surface (see Section *4c*).

### *3b.   Charge neutrality condition and the finite spatial footprint*

We would like to have the surface potential trap be electrostatic in nature so that it can be incorporated into the PNP equations without any problem. It would act as an externally applied field but with a small and finite footprint. Since we do not wish to dope the system with any electrical charges, the surface potential trap should not add or take away any charges from the system, i.e., it must be charge neutral. In addition, it should also have a limited spatial footprint as stated above. The latter is possible by considering the example of a capacitor with a positive charge layer separated from a negative charge layer with the same charge density. Outside the capacitor, there is no electrical field (or force) since it is overall electrically neutral. However, inside the capacitor there can be a very strong electric field. Our surface potential trap may be regarded as a generalization of this picture. It is also important to note that although in the following we specify a surface potential trap in the cylindrical geometry, the basic character of the surface potential trap, i.e., charge neutrality with a finite spatial footprint, is independent of the geometry, even though its form can change in accordance with geometric requirements.

The charge-neutral surface potential trap can be either positive or negative, depending on the physical properties of the fluid-solid interface. In the case of the silica-water interface, the surface potential should be positive in order to trap negative ions. To implement the surface potential trap so as to break the symmetry between the positive and negative ions, let us consider the trap function $f(r)$, where $r$ is the radial coordinate, that has two parameters—the height of the trap $\gamma$ and its width $\Delta$:

$$f(r) = \frac{\gamma}{2}\left(1 + \cos\frac{\pi(r-a)}{\Delta}\right), \quad \text{for} \quad a - \Delta \leq r \leq a \tag{3a}$$

$$f(r) = 0 \quad \text{for} \quad 0 \leq r \leq a - \Delta. \tag{3b}$$

The width of the surface potential trap, $\Delta$, is set to be the length of a hydrogen bond, about 8 Å. To verify that the functional form of $f(r)$ represents a charge neutral potential trap, we note that it must be related, through the Poisson equation, to a fixed underlying net charge density $\rho_C$, whose volume integral should be zero. That is, since $f(r)$ must satisfy the Poisson equation

$$\frac{1}{r}\frac{\partial}{\partial r}\left[r\frac{\partial f(r)}{\partial r}\right] = -\frac{\rho_C}{\varepsilon}, \tag{4}$$

the integration of $\rho_C$ over the domain $a - \Delta \leq r \leq a$ should yield zero, i.e., the potential trap does not bring any external net charges into the system. It is easy to demonstrate that the form of $f$ given by Eqs. (3) satisfies this constraint. Since the surface potential trap is regarded as an externally applied field; the underlying $\rho_C$ is fixed and treated as external to the system.

*3c.    Necessity for retaining a finite width*

Since the surface potential trap's width is very thin—8 Angstroms, one may be tempted to approximate it by a delta function.   However, we shall see that the finite width plays a significant role since it allows the mobile ions in the liquid to diffuse into the surface potential trap when the concentration gradient is sufficiently large.  This is an important element in realizing the IEP under the high acidity condition.   In other words, the finite width of the surface potential trap allows the mechanism of diffusion to play a role.

## IV.    The charge conserved Poisson-Boltzmann equation

*4a.    Static limit of the PNP equations*

The PB distribution can be obtained from the PNP equations by setting $J_-, J_+ = 0$.  In that static limit, we have

$$\nabla n_- - \frac{e}{k_B T} n_- \nabla \varphi = 0 \;, \tag{5a}$$

$$\nabla n_+ + \frac{e}{k_B T} n_+ \nabla \varphi = 0 \;, \tag{5b}$$

They can be integrated to yield

$$n_- = \alpha \exp[+e\varphi / k_B T], \tag{6a}$$

$$n_+ = \beta \exp[-e\varphi / k_B T], \tag{6b}$$

where $\alpha, \beta$ are the integration constants.   By setting $\alpha = \beta = n^\infty$, one immediately obtains Eq. (1).   However, in the present case the overall charge neutrality condition in our computational domain dictates that

$$\alpha \int_V d\mathbf{x} \exp[+e\varphi/k_BT] = n^O V = \beta \int_V d\mathbf{x} \exp[-e\varphi/k_BT], \qquad (7)$$

where $V$ denotes the volume of the system. Hence we see that $\alpha \neq \beta$ in general, in contrast to the previous assumption ($\alpha = \beta = n^\infty$) that led to the PB equation. From Eq. (6), it follows that

$$n_- = n^O \frac{\exp[+e\varphi/k_BT]}{\frac{1}{V}\int_V d\mathbf{x} \exp[+e\varphi/k_BT]},$$

$$n_+ = n^O \frac{\exp[-e\varphi/k_BT]}{\frac{1}{V}\int_V d\mathbf{x} \exp[-e\varphi/k_BT]}.$$

By substituting the above expressions into the Poisson equation, we obtain the following integral-differential equation for a cylindrical channel with radius $a$:

$$\frac{1}{r}\frac{\partial}{\partial r}\left(r\frac{\partial \varphi}{\partial r}\right) = \frac{ea^2 n^O}{2\varepsilon}\left[\frac{\exp(e\varphi/k_BT)}{\int_0^a r\exp(e\varphi/k_BT)dr} - \frac{\exp(-e\varphi/k_BT)}{\int_0^a r\exp(-e\varphi/k_BT)dr}\right]. \qquad (8)$$

It is easily seen that in contrast to the PB equation, Eq. (8) preserves the PNP equations' characteristic of being independent from an additive constant potential. The spatial integral of the right hand side of Eq. (8) is seen to yield zero. It is also easily verified that any uniform Dirichlet or Neumann boundary condition, in the absence of the surface potential trap, will yield trivial solutions, as mentioned previously.

If in addition we denote the potential generated by the net ionic charge density on the right hand side as $\psi$, and take into account the surface potential trap ($f$ can be incorporated into the PNP equation as part of the electrostatic potential), then the following equation is obtained:

$$\frac{1}{r}\frac{\partial}{\partial r}\left(r\frac{\partial \psi}{\partial r}\right) = \frac{ea^2 n^O}{2\varepsilon}\left\{\frac{\exp[e(\psi+f)/(k_B T)]}{\int_0^a \exp[e(\psi+f)/(k_B T)]rdr} - \frac{\exp[-e(\psi+f)/(k_B T)]}{\int_0^a \exp[-e(\psi+f)/(k_B T)]rdr}\right\}. \quad (9)$$

The above is denoted charge conserving Poisson-Boltzmann (CCPB) equation in the cylindrical geometry, where $\psi(r) = \varphi(r) - f(r)$. From Eq. (3b), it is clear that $\psi = \varphi$ for $r < a - \Delta$; this fact will be used to advantage in the re-derivation of the PB equation from the CCPB equation.

*4b.  Surface dissociated charge density*

With the surface potential trap and the CCPB, it is important to include the surface dissociated charge density as part of the total ion density $n_\pm^O$. Since $n_\pm^O(\pi a^2 L) = n_\pm^\infty(\pi a^2 L) + \sigma_\pm(2\pi a L)$, where $L$ denotes the length of the liquid channel, we have

$$n_\pm^O = n_\pm^\infty + 2\frac{\sigma_\pm}{a}, \quad (10)$$

where $\sigma_\pm$ denotes the interfacial dissociated charge densities. For a positive surface potential trap, $\sigma_-$ resides predominantly inside the surface potential trap, whereas $\sigma_+$ is in the Debye layer. From the previous discussion, it is clear that $\sigma_+$ and $\sigma_-$ are the $H^+$ and $OH^-$ ions, respectively. A physical understanding of Eq. (10) can be given as follows. With the presence of a positive surface potential trap, a high concentration of $OH^-$ is captured inside the domain of $f$. However, since the bulk ion density $n^\infty$ is a given constant (the $H^+$ and $OH^-$ ions are governed, in addition, by the law of mass action (see below)), it follows that there must be an overall increase in the ion densities from that given by the bulk ion densities. This fact is expressed by Eq. (10).

We denote the surface charge density *S* as the net charge density inside the surface potential trap, integrated over the region $a-\Delta < r < a$. The surface charge density *S*, when multiplied by the circumferential area of the liquid channel must be exactly equal in magnitude, but opposite in sign, to the net charge in the Debye layer. It should be noted that *S* is not necessarily equal to $\sigma_-$ inside the trap, since the surface potential trap is permeable to the bulk ions, in the sense that the ions in the liquid can enter and leave the surface trap. In particular, $\sigma$ represents a quantity that is averaged over the whole sample, whereas *S* pertains only to the surface potential trap region. Such ion flows, however, depend on many factors that include the acidity, the salt concentration, the liquid channel width, etc.

*4c.     Mathematical treatment to exclude non-surface-specific ions from the trap*

In the silica-water system the potential value at the interface is determined by the activity of the ions which react with the silica surface, i.e., the $H^+$ and $OH^-$ ions. Hence an important element in the surface reactivity is the pH value of the solution. It is also a physical fact that the other ions, e.g., those from the added salts and acids, cannot form part of the surface charge layer. Of course, mathematically one can simply let the ions other than the $H^+$ and $OH^-$ to "not see" the surface potential trap *f*, by associating *f* only with the $H^+$ and $OH^-$ ions. However, this has proven to be insufficient since such treatment cannot prevent, for example, the $Na^+$ ions from occupying the same spatial domain as *f*. This is especially the case since the surface potential trap can capture a high density of negative charges, which will attract the positive ions (other than the $H^+$ ions, such as the $Na^+$ ions) through the electrostatic interaction that is mathematically ensured by the Poisson equation. Such "leakage" of un-wanted ions (e.g., $Na^+$ ions) into the spatial domain of the surface potential trap $f$ can be especially detrimental to the proper description of the isoelectronic point and its related properties. And it has to be emphasized that such "leakage" cannot be completely stopped by having different surface potential values for different ions, since the electrical interaction is strong and always present.

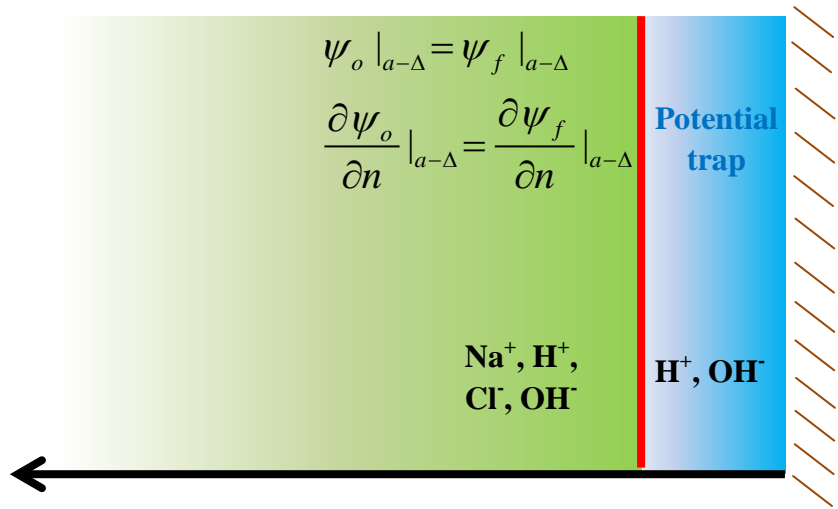

**Figure 2. A schematic illustration of the sub-domains in the solution of the CCPB equation, colored by green and blue. The solutions in the two regions are linked together by the two continuity conditions at the interface, denoted by the red line. This division of the computational domain of the Poisson equation is to ensure that no surface-non-specific salt or buffer ions can enter the surface potential trap. This physical condition is especially important in modeling the isoelectronic point and its related properties.**

In order to enforce mathematically the condition that *only* the $H^+$ and $OH^-$ ions can occupy the spatial domain of *f*, we divide the solution domain of the Poisson equation into two sub-domains as shown in Fig. 2. For the potential $\psi_o$ outside the trap (colored green) all the ion densities should be on the right hand side of the Poisson equation. For the potential inside the potential trap (colored blue), $\psi_f$, only the $H^+$ and $OH^-$ ion densities would appear on the right hand side of the Poisson equation. Solutions in the two sub-domains are then linked together by the two boundary conditions of the potential value and its normal derivative being continuous at $r = a - \Delta$, indicated by the red line in Fig. 2. This process will be made explicit in the next section, in conjunction with re-writing Eq. (9), which is nonlocal in character, into a local form via the definition of a global chemical potential $\mu$. It should be noted that an alternative approach to prevent the ions, other than the $H^+$ and $OH^-$ ions, to be in the vicinity of the interface is to have separate repulsive surface potential traps, $f_{Na}$ and $F_{Cl}$, for $Na^+$ and $Cl^-$ ions. However, this is not our choice in this work.

## V. Global chemical potential and the re-derivation of the Poisson-Boltzmann equation

*5a.    Re-writing the CCPB with the definition of a global chemical potential*

In the presence of the *NaCl* salt ions and/or *HCl* acid or the alkaline salt *NaOH*, we re-write Eq. (9) in the two regions, $r < a - \Delta$ and $a - \Delta < r < a$, respectively as

$$\frac{1}{r}\frac{\partial}{\partial r}\left(r\frac{\partial \psi_o}{\partial r}\right) = \frac{e}{\varepsilon}\{n_{Cl^-}^\infty \exp[e(\psi_o - \mu)/k_BT] + n_{OH^-}^\infty \exp[e(\psi_o - \mu)/k_BT]$$
$$- n_{Na^+}^\infty \exp[-e(\psi_o - \mu)/k_BT] - n_{H^+}^\infty \exp[-e(\psi_o - \mu)/k_BT]\} \quad , \quad (11a)$$

$$\frac{1}{r}\frac{\partial}{\partial r}\left(r\frac{\partial \psi_f}{\partial r}\right) = \frac{e}{\varepsilon}\{n_{OH^-}^\infty \exp[e(\psi_f - \mu + f)/k_BT] - n_{H^+}^\infty \exp[-e(\psi_f - \mu + f)/k_BT]\}.$$
(11b)

Here the dielectric constant $\varepsilon = \varepsilon_r \varepsilon_O$ with $\varepsilon_r = 80$ for water, $\varepsilon_O = 8.85 \times 10^{-12}$ F/m, and $n_{H^+}^\infty, n_{Na^+}^\infty, n_{Cl^-}^\infty$ and $n_{OH^-}^\infty$ are the bulk ion concentrations, with $n^{\infty}_{OH^-} + n^{\infty}_{Cl^-} = n^{\infty}_{Na^+} + n^{\infty}_{H^+}$. In the above $\mu$ is the global chemical potential, which arises from the overall charge neutrality constraint, i.e., the total integrated positive charges on the right hand sides of Eq. (11) should be equal to the total integrated negative charges:

$$\mu = \frac{k_B T}{2e} \ln \left\{ \frac{\int_0^{a-\Delta} \{n^{\infty}_{OH^-} \exp(e\psi_o/k_B T) + n^{\infty}_{Cl^-} \exp(e\psi_o/k_B T)\} r dr + \int_{a-\Delta}^{a} n^{\infty}_{OH^-} \exp[e(\psi_f + f)/k_B T] r dr}{\int_0^{a-\Delta} \{n^{\infty}_{H^+} \exp(-e\psi_o/k_B T) + n^{\infty}_{Na^+} \exp(-e\psi_o/k_B T)\} r dr + \int_{a-\Delta}^{a} n^{\infty}_{H^+} \exp[-e(\psi_f + f)/k_B T] r dr} \right\}.$$

(12)

It is to be noted that above definition of the global chemical potential is very similar to the approach used in semiconductor physics, with electrons and holes being the two types of charge carriers. In particular, it should be mentioned that the PNP equations have been extensively used in describing the physics of the PN junctions. Here the function of $\mu$ is to insure charge neutrality; and we distinguish it to be the *global* chemical potential, to be differentiated from the ion concentration and electrical potential, which form the two local components of the electrochemical potential and whose gradients give the two components of the ionic currents (see Eq. (2c)).

We should note that when Eqs. (11), (12) are considered together, an additive constant potential would just mean a constant shift of the solution, with no physical implications.

By solving Eqs. (11) and (12) simultaneously, one would obtain $\psi(x)$ and $\mu$, from which the total (average) ion density can be calculated as:

$$n^o = n^o_- = \frac{2}{a^2} \int_0^a \{n^{\infty}_{OH^-} \exp[e(\psi - \mu + f)/k_B T] + n^{\infty}_{Cl^-} \exp[e(\psi - \mu)/k_B T]\} r dr$$

$$= n^o_+ = \frac{2}{a^2} \int_0^a \{n^{\infty}_{H^+} \exp[-e(\psi - \mu + f)/k_B T] + n^{\infty}_{Na^+} \exp[-e(\psi - \mu)/k_B T]\} r dr$$

(13)

Since $n_{\pm}^{\infty}$ are the inputs to Eqs. (11) and (12), the knowledge of $n_{\pm}^{O}$ suffices to determine the interfacial dissociated charge densities $\sigma_{\pm}$ through Eq. (10). The values of $n_{H^+}^{\infty}$ and $n_{OH^-}^{\infty}$ are noted to be constrained by the law of mass action, $n_{H^+}^{\infty}(M) \bullet n_{OH^-}^{\infty}(M) = 10^{-14}(M^2)$, where $M$ denotes molar concentration. The law of mass action is noted to govern the equilibrium reaction rate, and in this case it is for the $H^+$ and $OH^-$ ions in acid or alkaline solutions.

*5b.    Re-derivation of the PB equation with its associated boundary value*

It should be especially noted that the PB equation can be re-derived from Eq. (11), but with an altered interpretation for its boundary value. By noting that $f(r) = 0$ for the reduced domain $r \leq a - \Delta$, Eq. (11a) may be written in the form

$$\frac{1}{r}\frac{\partial}{\partial r}\left(r\frac{\partial \psi}{\partial r}\right) = \frac{en^{\infty}}{\varepsilon}\left\{\exp[e(\psi - \mu)/k_B T] - \exp[-e(\psi - \mu)/k_B T]\right\}, \quad (14)$$

with $n^{\infty} = n_{Na^+}^{\infty} + n_{H^+}^{\infty} = n_{OH^-}^{\infty} + n_{Cl^-}^{\infty}$, within this *reduced domain*, which is noted to comprise only the mobile ions. Simple manipulation leads to the form of the PB equation:

$$\frac{1}{r}\frac{\partial}{\partial r}\left(r\frac{\partial \bar{\psi}^{(PB)}}{\partial r}\right) = \frac{1}{\lambda_D^2}\sinh\left(\bar{\psi}^{(PB)}\right). \quad (15)$$

Here $\bar{\psi}^{(PB)} = e(\psi - \mu)/k_B T$, with $\psi^{(PB)} = \psi - \mu$. The boundary condition, *applied at* $r = a - \Delta$, should be $\psi^{(PB)} = -\mu$ because we have set $\psi(a) = 0$, and therefore $\psi(a-\Delta) \to 0$ as $\Delta \to 0$ (in actual calculations, the difference from zero is at most a fraction of one mV, which is noted to be of the same magnitude of the traditional potential difference between the Stern layer and surface layer). It follows that in our form of the PB equation, $-\mu$ *plays the role of the traditional $\zeta$ potential* (apart from a very small potential difference across the surface potential trap). However, distinct from the traditional PB equation in which the $\zeta$ potential is treated as a constant, here $-\mu$ can vary with $n^\infty$ as well as other global geometric variations, such as the liquid channel radius (width). Since the use of Eq. (15) with the accompanying $-\mu$ boundary condition leads to exactly the same predictions as the CCPB equation, it is fair to say that the consideration of the charge neutrality constraint has led to a re-definition of the boundary condition for the PB equation.

*5c.    Definition of a generalized zeta potential from the Smoulochowski velocity*

In association with the above, we would also like to define a generalized zeta potential that can better reflect the electrokinetic activity in the nanochannels. Consider the application of an electric field $E_z = -\nabla g$ along the axial direction, denoted the $z$ direction, of the cylindrical channel to drive the liquid flow, by introducing a body force density in the Navier-Stokes (NS) equation. By using the Smoulochowski velocity expression, derived from the PB equation coupled with the NS equation, a clear relation between chemical potential $-\mu$ and zeta potential $\zeta$ can be obtained. We solve for the steady state solution under the condition that the ion density distribution profile along the cylindrical channel axial direction remains constant. The local electric field that arises from the ions can be ignored since it is perpendicular to the axial direction.

In the steady state, the velocity normal to axis is zero with $u_r = 0$. The axial velocity $u_z$ in the steady state can be written as

$$\frac{\partial^2 u_z}{\partial r^2} + \frac{1}{r}\frac{\partial u_z}{\partial r} = \frac{1}{\eta}\frac{\partial P}{\partial z} - \rho\frac{E_z}{\eta}, \tag{16}$$

where $\eta$ is the fluid viscosity and $P$ the pressure. The net ion charge density $\rho$ is related to electrical potential $\psi^{(PB)}$ via the Poisson equation:

$$\frac{\partial^2 \psi^{(PB)}}{\partial r^2} + \frac{1}{r}\frac{\partial \psi^{(PB)}}{\partial r} = -\frac{\rho}{\varepsilon}, \tag{17}$$

with $\varepsilon$ being the dielectric constant. Substituting the left hand side of Eq. (17) into Eq. (16) yields:

$$\frac{\partial^2 u_z}{\partial r^2} + \frac{1}{r}\frac{\partial u_z}{\partial r} = \frac{1}{\eta}\frac{\partial P}{\partial z} + \frac{\varepsilon E_z}{\eta}\left[\frac{\partial^2 \psi^{(PB)}}{\partial r^2} + \frac{1}{r}\frac{\partial \psi^{(PB)}}{\partial r}\right]. \tag{18}$$

The solution of Eq. (18), for $u_z$, can be expressed in terms of $\psi^{(PB)}$:

$$u_z = -\frac{\varepsilon E_z}{\eta}\left[(-\mu) - \psi^{(PB)}\right] + \frac{a^2 - r^2}{4\eta}\left(-\frac{dP}{dz}\right). \tag{19}$$

Here $-\mu$ is the boundary value of the potential and $a$ is the channel radius, but for the traditional PB model, the boundary value should be that at the infinite $a$ limit. The average axial velocity can be calculated, in terms of the solution potential profile:

$$\overline{u_z} = \frac{1}{a^2}\int_0^a 2u_z r dr = -\frac{2\varepsilon E_z}{\eta}\int_0^1 \left[(-\mu) - \psi^{(PB)}\right]\left(\frac{r}{a}\right)d\left(\frac{r}{a}\right) = -\frac{\varepsilon E_z}{\eta}\zeta, \tag{20}$$

with $dP/dz = 0$. It is seen that $\overline{u_z}$ is proportional to the $\zeta$ potential. Hence we would like to define the zeta potential from Eq. (20) as [19]:

$$\zeta = -\frac{2}{a^2}\left(\int_0^{a-\Delta}\psi_o(r)rdr + \int_{a-\Delta}^a \psi_f(r)rdr\right). \tag{21}$$

The zeta potential expresses the average potential drop between the liquid-solid interface and the center of the channel. It reflects the electrokinetic driving force for the system.

*5d. Solution procedure and the interfacial-related quantities*

Here we summarize the solution procedure of our approach. By using the package COMSOL Multiphysics version 4.4, one can solve Eqs. (11) and (12) simultaneously in a self-consistent manner, with two sub-domains as shown in Fig. 2. The boundary conditions used are $\psi_f |_{r=a} = 0$, $\frac{\partial \psi_o}{\partial r}|_{r=0} = 0$. The pH value, salt concentration, law of mass action, and charge neutrality constraints determine the inputs $n_\pm^\infty$ for all the ions. The outputs are the potential $\psi(x)$ plus the chemical potential $\mu$. From Eq. (13) we then obtain $n_\pm^O$ and from Eq. (10) the interfacial dissociation charge density $\sigma_\pm$. The (net) surface charge density $S$ in the potential trap can be obtained as:

$$S = \frac{1}{a}\left( n_{H^+}^\infty \int_{a-\Delta}^{a} \exp[-e(\psi_f - \mu + f)/k_B T] r dr - n_{OH^-}^\infty \int_{a-\Delta}^{a} \exp[e(\psi_f - \mu + f)/k_B T] r dr \right) \quad (22)$$

Here $S$ represents the surface charge density that should exactly cancel the net charge in the diffuse Debye screening layer. It is the total charge in the Stern layer. A difference between $S$ and $\sigma = \sigma_-$ in the trap is seen in Fig. 3(a), which is due to the fact that whereas $S$ is the net charge inside the surface potential trap, $\sigma$ represents the globally averaged value. Since there is a deficit of $OH^-$ ions in the Debye layer, in the vicinity of the surface potential trap region, hence when averaged over the sample volume we always have $S > \sigma$. However, when the radius decreases below $\lambda_D$, $S$ is seen to approach $\sigma$. The fact that the surface charge density $S$ decreases with decreasing channel width is generally denoted as a manifestation of the "charge regulation" phenomenon. Under very acid environment, it will be seen below that the value of $S$ can approach zero and even become positive, a phenomenon denoted as the "isoelectronic point," owing to the diffusion of the $H^+$ ions into the surface potential trap (caused by the extremely large concentration gradient between the outside and inside the potential trap). In the holistic approach these phenomena are seen to appear naturally, as the consequence of the static limit of the PNP equations and the global charge neutrality constraint in the presence of a surface potential trap.

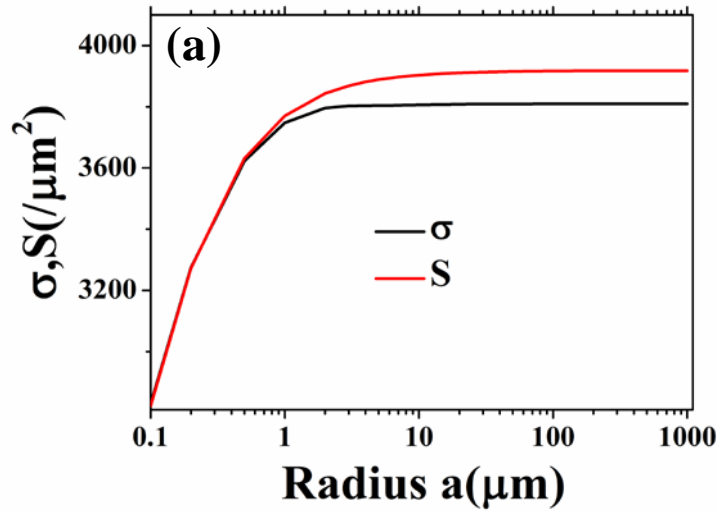

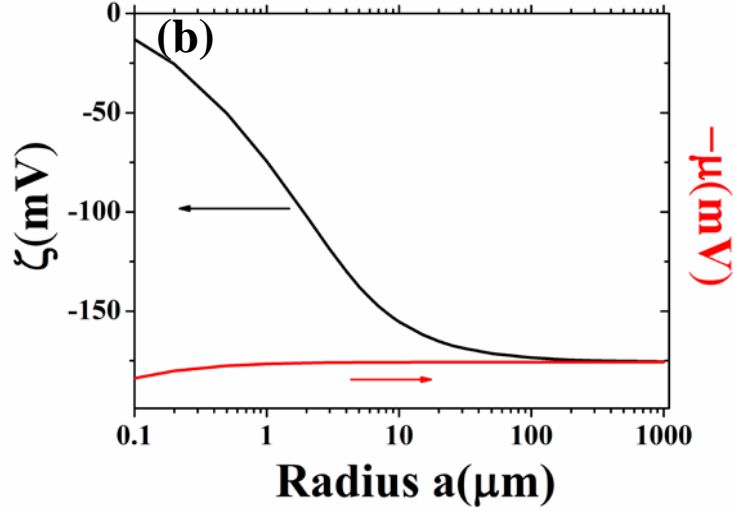

**Figure 3.** The self-consistently determined interfacial dissociated charge density $\sigma = \sigma_- = \sigma_+$ as defined by Eq. (10), plotted as a function of $a$ (black curve). The red curve is for $S$, defined as the density of the ions integrated over the width of the surface potential trap. It is seen that $S > \sigma$ because part of $S$ is captured from the bulk. (b) Negative of the chemical potential, $-\mu$ (right scale, red curve) plotted as a function of $a$. The black curve denotes the zeta potential, $\zeta$ (left scale), as defined by Eq. (21). It is seen that the two quantities agree closely in the large $a$ limit, but deviate from each other when $a$ decreases. The calculated case is for pH7, with no salt added. The energy height of the potential trap used is $\gamma=510$ mV.

In Fig. 3(b) we show the associated variation of $-\mu$ plotted as a function of *a*, where it is seen that $\zeta$ has the same value as $-\mu$ in the large channel limit, but the two deviate from each other as the liquid channel width diminishes.

In all our numerical calculations presented in this work there is only one adjustable parameter, the height of the surface potential trap $\gamma$=510 mV. The width of the potential trap is fixed at $\Delta$=8 Angstroms.

## VI. Predictions of the holistic approach

Owing to the inclusion of the surface charge layer as part of the computational domain in the holistic approach, it becomes possible to evaluate various parameters and predict some observed phenomena that are previously beyond the traditional PB equation alone.

*6a.    Surface capacitance and surface reactivity*

We first evaluate the surface capacitance and surface reactivity constants, denoted the pK and pL values, that were traditionally assumed to be obtainable only with the help of experimental inputs [9,11-13].

As counter ions dissociate from the surface, they form a diffuse cloud of mobile charges within the electrolyte. The Stern layer model treats the counter ions as being separated from the surface by a thin Stern layer across which the electrostatic potential drops linearly from its surface value $\psi_0$ to a value $\psi_d$, called the diffuse layer potential. This potential drop is characterized by the Stern layer's phenomenological capacitance, $C = \dfrac{S}{\psi_0 - \psi_d}$. This capacitance reflects the structure of silica-water interface and should vary little with changes in surface geometry or electrolyte concentration. We calculate the capacitance using $C = \dfrac{S}{\Delta\psi_f}$, where $\Delta\psi_f$ means the potential drop inside the potential trap. In our calculation the value of $\Delta\psi_f$, a small but nonzero quantity, is easily obtained, so is $S$. The calculated capacitance, in pH range of 5 to 9, is around 1.3 F/m$^2$. This value is noted to lie within the range of reported values that can vary from 0.2 to 2.9 F/m$^2$ over the same pH range [35].

The two reactions that can happen on the silica/water interface are: $SiO^- + H^+ \Leftrightarrow SiOH$, and $SiOH + H^+ \Leftrightarrow SiOH_2^+$. The latter is significant only under high acidity conditions. The equilibrium constants of these two reactions are defined by

$$K = \frac{N_{SiO^-}[H^+]_o}{N_{SiOH}} = 10^{-pK}\,[mol/L],$$

and

$$L = \frac{N_{SiOH}[H^+]_o}{N_{SiOH_2^+}} = 10^{-pL}\,[mol/L].$$

Here $[H^+]_o$, in units of [$mol/L$], is the proton local density at the outer boundary of the surface potential trap; and $N_{SiO^-}$, $N_{SiOH_2^+}$, and $N_{SiOH}$, all in the same unit of [nm$^{-2}$], are the surface densities of the respective $SiO^-$, $SiOH_2^+$, and $SiOH$ groups. For $N_{SiO^-}$ we simply use the negative ion density (that of $OH^-$) inside the potential trap, integrated over its width $\Delta$ to yield the surface density. Here we are reminded of the reaction $SiOH + OH^- \Leftrightarrow SiO^- + H_2O$ (see Section *3a*), so that the surface density of $OH^-$ is treated the same as that of $SiO^-$. The value of $[H^+]_o$ can be simply obtained from our calculation at the position just outside the surface potential trap. For the $N_{SiOH}$, one can use the total site density, $1/v_o$, where $v_o$ denotes the average volume occupied by a single silicon dioxide molecule, and approximate $N_{SiOH} \approx 1/v_o^{2/3} = (0.35^3)^{-2/3}$ nm$^{-2}$ = 8.2 nm$^{-2}$. This value is noted to be very close to a commonly cited literature value for nonporous, fully hydrated silica, $N_{SiOH}$ = 8 nm$^{-2}$ [9]. The pK value so obtained is in the range of 7.14-7.28 for the pH range of 3 to 10 as shown in Fig.4. The pK value, usually considered to be independent of salt concentration and pH values (5-9), turns out to display some variation when the pH value or salt concentration increases. This agrees reasonably well with the literature reported pK values that can range from 4 to 6-8 [36] within the same pH range.

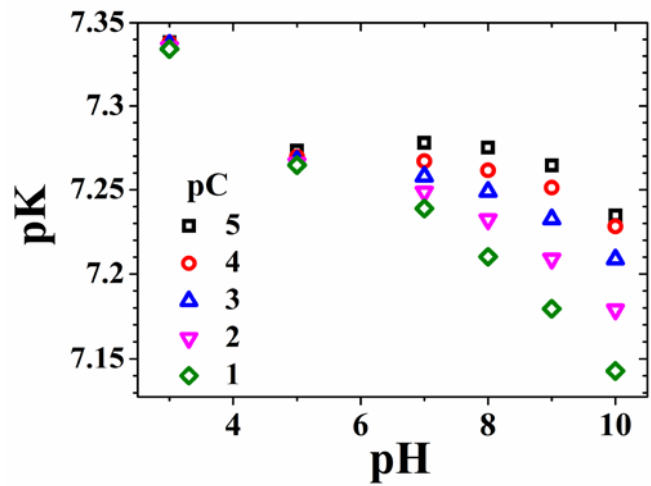

**Figure 4.** The pK values obtained from definition of the equilibrium constant of the **reaction** $SiO^- + H^+ \Leftrightarrow SiOH$. **The energy height of the potential trap used is γ=510 mV.**

For the second reaction that can occur under high acidity conditions, we shall take $N_{SiOH_2^+}$ as the positive surface charge density in the potential trap. At pH2, the derived pL value is $-2.23$. This is again in rough agreement with the reported pL values, which can range from $-3.5$ to $-1$, or 3 to 4 [36].

*6b.     Isoelectronic point and related properties in its vicinity*

In this subsection we show that the holistic approach can satisfactorily explain the appearance of the isoelectronic phenomenon and its related behaviors with just one adjustable parameter, i.e., the height of the surface potential trap $\gamma$, set at 510 mV. Experimentally, the IEP value has been observed to be in the range of pH2.5 to pH3.2 [37], i.e., under the high acidity condition.  Physically, one expects that under such conditions the proton concentration is so high that a fraction of the $H^+$ ions can be driven into the surface layer by the huge concentration gradient (in spite of the unfavorable energy consideration), so as to neutralize the surface charge density. Thus in modeling this phenomenon it is appreciated that the finite width of the surface potential trap can play an important role.

In Fig. 5(a) we compare the theory prediction of the surface charge densities (calculated for a large channel radius of 20 μm) to that measured from silica particles

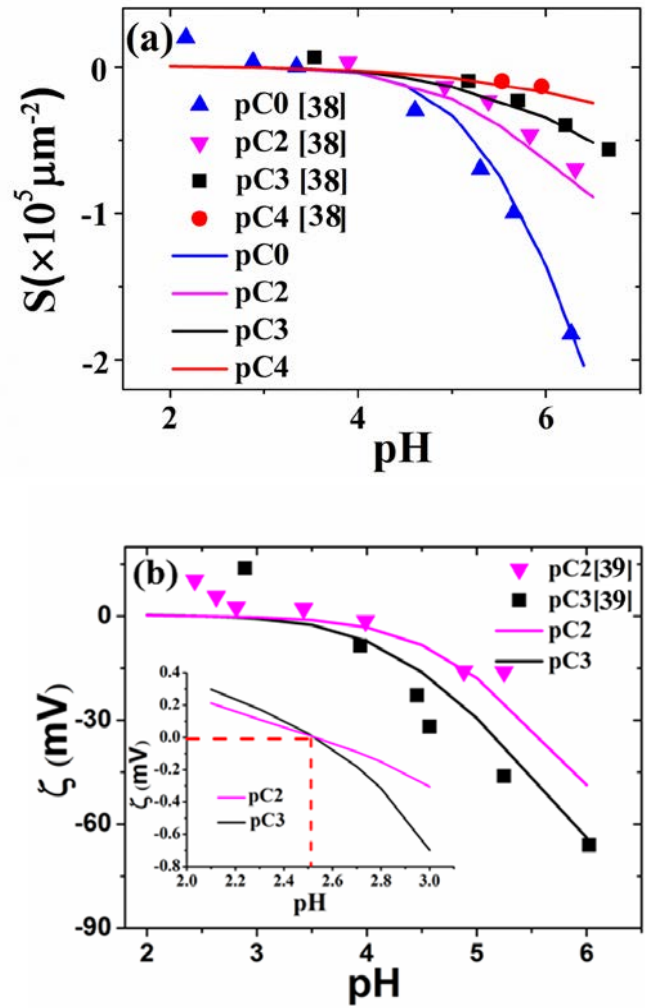

**Figure 5.** (a) Calculated surface charge density plotted as a function of pH values with a channel radius of 20 μm (solid lines). Experimental data [38] are shown as filled symbols. Excellent theory-experiment agreement is seen. (b) Zeta potential plotted as a function of pH values under different salt concentrations, with a channel radius of 20 μm. Theory predictions are shown as the solid lines, and experimental data are shown as filled symbols [39]. Semi-quantitative agreement is seen. Inset shows an enlarge view of zeta potential around pH 2-3. The zeta potential is seen to cross zero at the same isoelectronic point, pH 2.5, for two different salt concentrations. All the solid curves were calculated with γ=510 mV.

[38], both plotted as a function of pH, for various salt concentrations (i.e., pC values). Excellent quantitative agreement is seen. For high salt concentrations, the variation of the surface charge density as a function of the pH values is seen to be sharper. In other words, the screening effect of the salt ions is seen to enhance interfacial charge

separation. The zeta potential is found to vanish at pH2.5, consistent with the experimental measurements that indicate the IEP to be around pH2.5 to pH3.2. Figure 5(b) shows that the theory prediction of the zeta potential displays similar trends and magnitudes, in semi-quantitative agreement. The magnitude of the zeta potential is seen to increase as the salt concentration decreases. This is attributed to the fact that at lower salt concentrations (larger pC values), the screening effect is less prominent. The inset to Fig. 5(b) shows that the value of IEP is insensitive to the salt concentration.

Associated with the IEP is the well-known phenomenon of electrical double layer inversion. In a very acid environment, such as that close to the IEP, Debye length becomes comparable to the potential trap thickness and loses its usual implications. In contrast to the situation near pH7 in which one expects an accumulation of protons near the interface that results from charge separation, in a very acid environment the proton concentration can actually see a depletion at the interface.

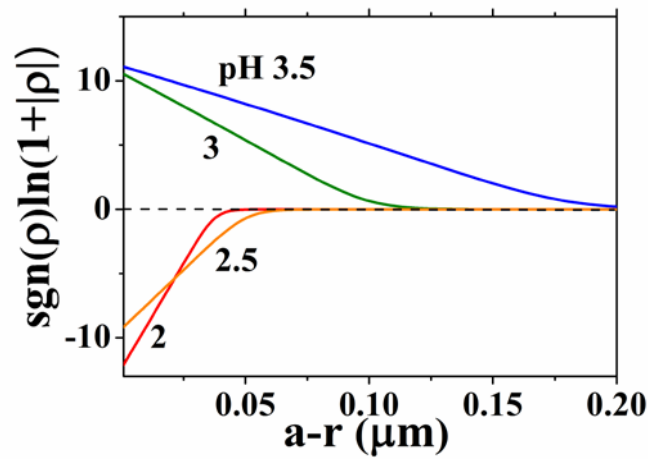

**Figure 6. Spatial distribution of local net charge concentration under different pH**

**values outside the potential trap, where ρ=$n_+ - n_-$ is in units of μm$^{-3}$. The surface charge layer, not resolved here, must have the opposite sign as compared to the diffuse layer so as to maintain charge neutrality. Hence a clear inversion in the electrical double layer is seen between pH3 and pH2.5. All results were calculated with *γ*=510mV.**

Hence if one lets $n_{H^+}^O$ to denote the total $H^+$ ion density for the system, then $n_{H^+}^O > n_{H^+}^\infty$ for large pH. However, the reverse situation, $n_{H^+}^O < n_{H^+}^\infty$, occurs close to the IEP. Associated with this is the electrical double layer inversion as illustrated in Fig. 6, which shows that between pH3 and pH2.5 there is clearly an inversion. It is interesting to note that the net polar orientation of interfacial water molecules was observed to flip close to pH4 [40].

*6c.    Broken geometric symmetry and the appearance of the Donnan potential in nanochannels*

In addition to the pH environment, geometry and size of the systems also play an important role in electrokinetics. Extended nanofluidics, the study of fluidic transport at the channel size on the order of 10-1000 nm, has emerged recently in the footsteps of microfluidics [41]. In almost all the applications it is also usually the case that the nanochannels are embedded in a large reservoir, so that there is no longer the geometric symmetry shown in Fig. 1. It is important to note that the small size of the nanofluidic channels allows many unique applications [42,43]. But it is precisely in such nanofluidic channels that the traditional approach, based on the PB equation, fails to give an accurate and detailed description of the physical situation owing to the fact that the characteristic dimension of the channel width is comparable to, or smaller than, the Debye length, so that the surface charges are significantly influenced by the liquid ionic distribution, and vice versa (hence both can vary along the interface). It is to be noted that such surface effects have enabled unique chemical operations, such as ion concentration [44] and rectification [45].

We show that the same theoretical framework can be applied to obtain the Donnan potential of a nanochannel in equilibrium with a large reservoir, i.e., when the geometric symmetry shown in Fig. 1 is broken.  In particular, it shows that the Donnan potential arises from the electrical double layer at the inlet regions of the nanochannel, and such double layer would disappear when the channel radius is large so that that the Debye layers on opposite sides of the channel do not overlap. Conversely, the Donnan potential increases with decreasing nanochannel radius so that the Debye layers overlap each other.

Behaviors in confined spaces can differ from those in the bulk even when they are linked to each other. To take account of the equilibrium between the bulk and the confined space, we consider an extended nanochannel bridging two large chambers, here denoted as the "reservoir."  The extended nanochannel has a radius of 0.2 μm and a height $h$ of 0.4 μm.  The reservoir has a radius of 0.7 μm and height of 7.85 μm, one on each side. They are partially shown in Fig. 7(a). Boundary condition at the cylindrical wall of the reservoir is defined to have inversion symmetry about its axis. Zero normal flux is applied at the reservoir's upper (and lower) boundary.  At the mid-plane of the bridging channel the reflection symmetry boundary condition is applied.  The narrow channel is confined by silica sidewalls with a relative dielectric constant $\varepsilon_r^s = 4$. The usual electrostatic boundary conditions are applied at the silica wall, which also has a surface potential trap with $\gamma$=510mV. However, the side of the silica facing the reservoir is considered to be coated with a thin layer of surface inactive material and hence no surface potential exists.

The calculated results for pH6.22, with no added salt, are displayed in color in Fig. 7(a). The left panel of 6(a) shows the net charge (in units of electronic charge) per unit length, obtained by integrating the charge density over each cross-section. It is seen that an electrical double layers is established at the inlet region of the nanochannel, with the (positive) net charge on the reservoir side decaying to zero in about 3 microns away from the nanochannel inlet. This electrical double layer is responsible for the Donnan potential [46] of the nanochannel, shown in Fig. 7(b). The Donnan potential saturates after a certain reservoir heights. In this case $V_D$ remains unchanged when reservoir height exceeds 6 μm, with a value of $V_D^{\infty}=-130$ mV. It should be noted that the electrical force density (on the charges) in the double-layer region is accurately counter-balanced by the osmotic pressure gradient given by the van't Hoff formula, $\Pi = k_B T \nabla n_+$ [47] as shown in Fig. 7(c), so that the equilibrium is attained.

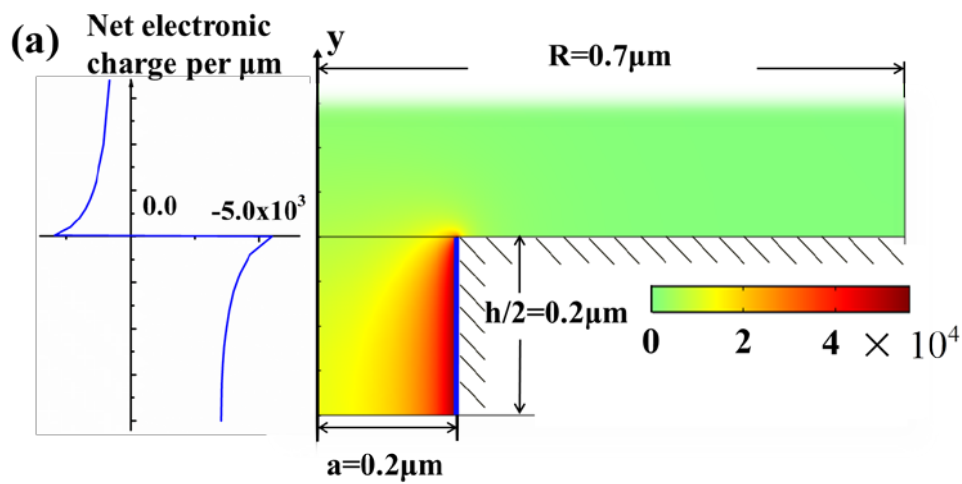

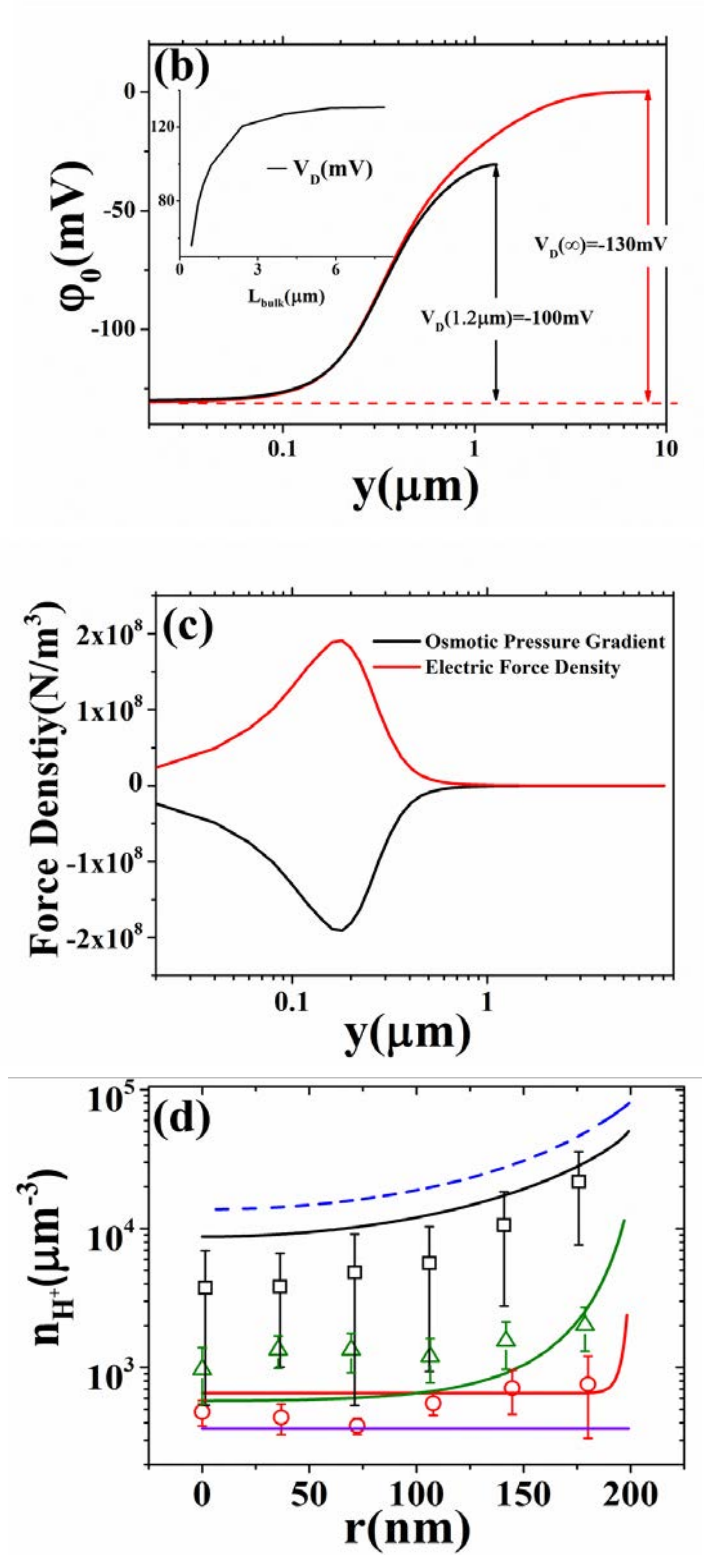

**Figure 7.** (a) Net charge concentration (shown in color, in unit of number of electronic charges per μm$^3$) in a channel with radius of 0.2 μm that is in contact with a reservoir (shown partially, the reservoir height is 7.85μm). Left panel shows the net electronic charge density integrated over the cross section (blue line), plotted along the y-axis (shown partially). The net charge is positive on the reservoir side and negative inside the nanochannel, thereby forming an electrical double layer. The pH value in the reservoir is set at pH6.22 (no salt addition) so as to agree with the experimental value [48]. The shaded region is the silica with a dielectric constant $\varepsilon_r^s =4$. (b) Electrical potential $\varphi_o$ plotted along the axis of the cylindrically shaped computational domain. The black line stands for reservoir height of 1.2 µm and the red line stands for reservoir height of 7.85 µm. The latter represents the plateau value as the reservoir height increases towards infinity. Donnan potential, $V_D$, represents the potential difference between the nanochannel and the reservoir. It clearly arises from the electrical double layer established at the inlet region of the nanochannel. Inset: $V_D$ increases as reservoir height increases, and reaches a plateau value around 6 µm. (c) Osmotic pressure gradient and electrical force density for the case where the reservoir height is 7.85µm. Very accurate counter-balance is seen between the two, as it should. (d) Cross sectional proton concentration distribution, averaged over the length of the channel. Black open squares with error bars are experimental data from reference [48] with a channel width of 410 nm and pH 6.22 (no salt addition). The black line is the corresponding theory prediction with the same experimental parameters. The blue dashed line represents the theory prediction in the absence of the reservoir. The green open triangles are experimental data from reference [48] with pH 6.03 and 0.0001M salt concentration; the solid green line is the corresponding theory prediction. The red open circles are data from reference [48] with pH 5.92 and 0.01M salt concentration; the solid red line is the corresponding theory prediction. The solid magenta line is the reference bulk proton density at pH6.22.

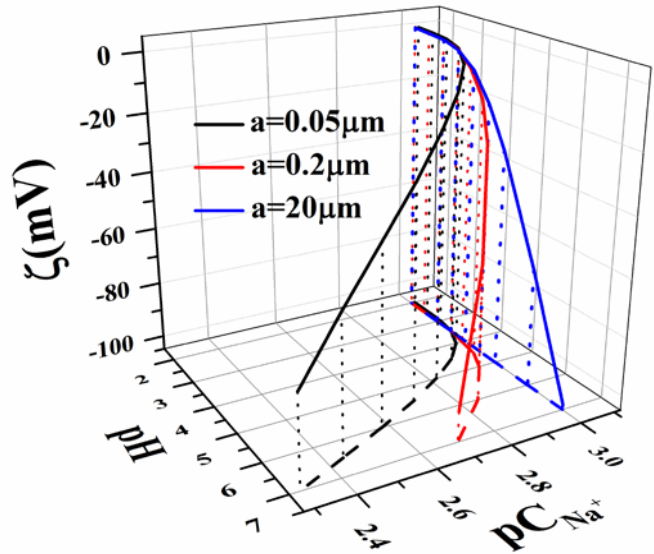

**Figure 8.** Zeta potential plotted as a function of bulk pH value and negative logarithm of Na$^+$ concentration, pC$_{Na^+}$ inside the nanochannel, for a set of channels with different radii (in µm). The salt concentration is 1mM. All results were calculated with γ=510mV.

Owing to the short length of the nanochannel, the decay of the net charge at the inlet can extend to the entire nanochannel. As a result, a clear enhancement in proton concentration can be seen in the extended nanospace. Here we model a case with geometric dimensions and other relevant parameters taken from the experiment of Kazoe et al. [48]. In Fig. 7(d), the black line denotes the theory prediction for the cross sectional proton distribution, averaged over the length of the channel. This is seen to be consistent with the experimental observation of Kazoe et al. [48] as shown by open symbols. Here the dashed blue line represents the model prediction in the absence of a reservoir. There is a clear enhancement of the proton density in the confined space when compared to that in the bulk (solid magenta line). With the addition of salt, the proton concentration is lowered in the confined nanospace (green and red solid lines), in agreement with the experimental observations (green and red symbols).

If salt is added, the positive salt ions will also show increased concentration inside the extended nanochannels. For the pH<7 case, we have calculated the zeta potential inside two nanochannels, radii of 0.2 μm and 0.05 μm, that are in equilibrium with a reservoir which has the same dimensions as that shown in Fig. 7(a). In addition, we have also calculated the reference case in which the channel radius is 20 μm. In Fig. 8 the results are plotted as a function of both pH and the average $Na^+$ ions' concentration inside the channel. The purpose here is to illustrate the effect of the channel radius on both the zeta potential as well as the $Na^+$ ions concentration when a fixed 1 mM of bulk salt concentration is added.

It is important to note that as the nanochannel's length increases, the net charge density at the central cross section of the channel approaches zero. Hence the net charge is an effect introduced by the broken geometric symmetry of the system. Also, as the channel radius increases so that the Debye layers on the opposite sides of the channel wall no longer overlap each other, the charging effect at the inlet regions disappears, and the Donnan potential $V_D$ approaches zero. Conversely, decrease in the nanochannel radius increases the Donnan potential magnitude. In particular, $V_D^\infty = -167.2 mV$ and $-154 mV$ for nanochannel radii of 0.05 μm and 0.1 μm, respectively. Hence there is a clear nanochannel radius dependence of the inlet charging effect and the associated Donnan potential.

Due to the rapid variation in the ionic densities in the inlet region of the nanochannel, we have also observed that the magnitude of the surface charge density can vary as well, generally in the range of a 2-3% increase. Such an effect is in the nature of the "charge regulation" phenomenon, but in this case it occurs because of the geometric symmetry breaking.

## VII. Concluding remarks

In conclusion, we show that the holistic approach to the charge separation phenomenon at the water-silica interface, based on the consideration of electrical energetics, can predict a plurality of observed physical effects that are beyond the traditional PB equation alone. Our approach is based on the PNP equations, with the charge seperation process driven by the introduction of a charge-neutral surface potential trap. The surface charge layer and the Debye layer are consistently considered within a single computational domain. The PB equation is re-derived within our formalism with a new interpretation for its boundary value. By using a single value of the phenomenological parameter which is the height of the surface potential trap, our approach is shown to yield predictions of surface capacitance, the pK and pL values, the isoelectronic point with its related phenomena, and the appearance of the Donnan potential in nanochannels, among others. All these predictions are shown to be in good agreement with the experimental observations. The holistic approach offers conceptual and computational simplicity in obtaining the information regarding the interfacial charge separation phenomena involving fluid with varying acidity (alkalinity) and salt concentrations, as well as channels of various width and broken geometric symmetry. It is capable of dealing with problems involving interfaces with complex geometries, which can be much more difficult by using the traditional approach.

P.S. wishes to acknowledge the support of SRFI11/SC02 and RGC Grant HKUST604211 for this work.